\documentclass[aps,prd,floats,twocolumn,twoside,preprintnumbers,
superscriptaddress,floatfix,nofootinbib,showpacs]{revtex4-1}
\usepackage{latexsym,bbold}
\usepackage{bm}

\usepackage[fleqn]{amsmath}
\usepackage{amssymb}

\usepackage{color}

\usepackage{graphicx}

\newcommand{\eqnb}{\begin{equation}}
\newcommand{\eqne}{\end{equation}}

\begin{document}


\let\I\i
\def\i{\mathrm{i}}
\def\e{\mathrm{e}}
\def\d{\mathrm{d}}
\def\half{{\textstyle{1\over2}}}
\def\thalf{{\textstyle{3\over2}}}
\def\h{{\scriptscriptstyle{1\over2}}}
\def\th{{\scriptscriptstyle{3\over2}}}
\def\fh{{\scriptscriptstyle{5\over2}}}
\def\vec#1{\mbox{\boldmath$#1$}}
\def\svec#1{\mbox{{\scriptsize \boldmath$#1$}}}
\def\oN{\overline{N}}
\def\ttimes{{\scriptstyle \times}}
\def\bm#1{{\pmb{\mbox{${#1}$}}}}

\def\CG#1#2#3#4#5#6{C^{#5#6}_{#1#2#3#4}}
\def\threej#1#2#3#4#5#6{\left(\begin{array}{ccc}
    #1&#2&#3\\#4&#5&#6\end{array}\right)}

\title{$\eta'$ and $\eta$ mesons at high $T$ when the $U_A(1)$ and
 chiral symmetry breaking are tied}

\author{Davor Horvati\' c}
\affiliation{Physics Department, Faculty of Science,
University of Zagreb, Bijeni\v{c}ka cesta 32, 10000 Zagreb, Croatia}
\author{Dalibor Kekez}
\affiliation{Rugjer Bo\v{s}kovi\'{c} Institute, Bijeni\v{c}ka cesta 34,
10000 Zagreb, Croatia}
\author{Dubravko Klabu\v{c}ar}
\affiliation{Physics Department, Faculty of Science,
University of Zagreb, Bijeni\v{c}ka cesta 32, 10000 Zagreb, Croatia}

\date{\today}

\begin{abstract}
The approach to the $\eta'$-$\eta$ complex employing chirally well-behaved 
quark-antiquark bound states and incorporating the non-Abelian axial anomaly
of QCD through the generalization of the Witten-Veneziano relation, is extended 
to finite temperatures. Employing the chiral condensate has led to a sharp chiral
and $U_A(1)$ symmetry restoration; but with the condensates of quarks with
realistic explicit chiral symmetry breaking, which exhibit a smooth, crossover
chiral symmetry restoration in qualitative agreement with lattice QCD results,
we get a crossover $U_A(1)$ transition, with smooth and gradual melting of
anomalous mass contributions. This way we obtain a substantial drop of the
$\eta'$ mass around the chiral transition temperature, but no $\eta$ mass drop.
 This is consistent with present empirical evidence.
\end{abstract}


\maketitle

\section{Introduction}
\label{INTRO}

The experiments on heavy-ion collider facilities, such as RHIC, LHC, FAIR, and NICA, aim
to produce a new form of hot and/or dense QCD matter \cite{Akiba:2015jwa,Dainese:2016dea}.
Clear signatures of its production are thus very much needed.
The most compelling such signal would be a
change of a pertinent symmetry, {\it i.e.}, restoration - in hot and/or
dense matter - of the symmetries of the QCD Lagrangian which are broken
in the vacuum, notably the [$SU_A(N_f)$ flavor] chiral symmetry for 
$N_f=3=2+1$ light quark flavors $q$, and the $U_A(1)$ symmetry. This provides 
a lot of motivation to establish that experiment indeed shows it, as well 
as to give theoretical explanations of such phenomena.


The first signs of a (partial) restoration of the $U_A(1)$ symmetry were claimed
to be seen in 
 200 GeV Au+Au collisions \cite{Adler:2004rq,Adams:2004yc} 
at RHIC by Cs\"org\H{o} {\it et al.}  \cite{CsorgoVertesiSziklaiVargyas}.
They analyzed the $\eta'$-meson data of PHENIX \cite{Adler:2004rq} and STAR
\cite{Adams:2004yc} collaborations through several models for hadron multiplicities,
and found that the $\eta'$ mass ($M_{\eta'}=957.8$ MeV in the vacuum) drops by at
least 200 MeV inside the fireball. The vacuum $\eta'$ is, comparatively, so very
massive since it is predominantly the $SU_V(N_f)$-flavor singlet state $\eta_0$.
Its mass $M_{\eta_0}$ receives a sizable anomalous contribution $\Delta M_{\eta_0}$
due to the $U_A(1)$ symmetry violation by the non-Abelian axial Adler-Bell-Jackiw
 anomaly (`gluon anomaly' or `$U_A(1)$ anomaly' for short),    
which makes the divergence of the singlet axial quark current
${\bar q}\gamma^\mu \gamma_5 {\frac{1}{2}} \lambda^0 q$ nonvanishing even in the 
chiral limit of vanishing current masses of quarks, $m_q\to 0$. 
The said mass drop is then the sign of a {\it partial} $U_A(1)$ symmetry restoration
 in the sense of diminishing contribution of $U_A(1)$ anomaly to the
 $\eta'$ mass, which would drop to a value readily understood in the same way 
 \cite{Kapusta:1995ww} as the masses of the octet of the light pseudoscalar
 mesons $P=\pi^{0,\pm}, K^{0,\pm}, {\bar{K^0}}, \eta$, which are exceptionally
 light almost-Goldstone bosons of Dynamical Chiral Symmetry Breaking (DChSB). 

Now, there is a new experimental paper \cite{Adare:2017vig}
on 200 GeV Au+Au collisions.
 Although a new analysis of the limits on $\eta'$ and $\eta$ masses was
 beyond the scope of Ref. \cite{Adare:2017vig}, the data contained therein
 make it possible, and preliminary considerations
 \cite{CsorgoMexicoSymposiumSept2017} 
 confirm the findings of Refs. \cite{CsorgoVertesiSziklaiVargyas}.


The first explanation \cite{Benic:2011fv} of these original findings
 \cite{CsorgoVertesiSziklaiVargyas}
was offered by conjecturing that the Yang-Mills (YM) topological susceptibility,
which leads to the anomalously high $\eta'$ mass, should be viewed through the 
Leutwyler-Smilga (LS) \cite{Leutwyler:1992yt} relation (\ref{LS}). This 
ultimately implies that the anomalous part of the $\eta'$ mass falls 
together with the quark-antiquark ($q\bar q$) chiral-limit condensate 
$\langle{\bar q}q\rangle_0(T)$ as the temperature $T$ grows towards the 
chiral restoration temperature $T_{\rm Ch}$ and beyond.
This tying the $U_A(1)$ symmetry restoration with the chiral symmetry one,
was just a conjecture until our more recent paper \cite{Benic:2014mha}
strengthened the support for this scenario. Nevertheless, there was also a 
weakness: our approach predicted the drop of not only the $\eta'$ mass, but 
also (even more drastically) of the $\eta$ mass $M_{\eta}$, and signs for
that have not been seen in any data, including the new \cite{Adare:2017vig}
and the newest \cite{Aidala:2018ond}. 
In the present paper, we show that the predicted  \cite{Benic:2011fv}
drop of $M_{\eta}$ was the consequence
of employing the chiral-limit condensate $\langle{\bar q}q\rangle_0(T)$, 
since it falls too fast with $T$ after approaching $T \sim T_{\rm Ch}$.
We then perform $T>0$ calculations in the framework of the more recent
Beni\'c {\it et al.} \cite{Benic:2014mha}, where  LS relation (\ref{LS})
is replaced by the full QCD topological charge parameter (\ref{defA}) 
\cite{Shore:2006mm,Shore:2x,Di Vecchia:1980ve}.
There one can employ $q\bar q$ condensates for realistically massive
 $u, d$ and $s$-quarks, with much smoother $T$-dependence.
As a result, the description of the $\eta$-$\eta'$ complex of Ref.
\cite{Benic:2011fv} is significantly improved, since our new $T$-dependences
of the pseudoscalar meson masses do not exhibit a drop of the $\eta$ mass,
while a considerable drop of the $\eta'$ mass still exists, consistently
with the empirical findings \cite{CsorgoVertesiSziklaiVargyas}.

\section{A survey of the $\eta$-$\eta^\prime$ complex}
\label{survey}

\noindent The light pseudoscalar mesons are {\it both} $q{\bar q'}$
bound states ($q,q'=u,d,s$), and, simultaneously, 
(almost-)Goldstone bosons of DChSB of nonperturbative QCD. The approach which
 simultaneously implements {\it both}, is the one through the Dyson-Schwinger 
(DS) equations for Green functions of QCD. (See, {\it e.g.,} Refs.
\cite{Alkofer:2000wg,Roberts:2000aa,Holl:2006ni,Fischer:2006ub}
 for reviews.)
Presently pertinent is the gap equation for {\it dressed} quark propagators 
$S_q(p)$ with DChSB-generated self-energies $\Sigma_q(p)$:
\begin{equation}
{S}_q^{-1}(p) = S^{\scriptstyle {\rm free}}_{q}(p)^{-1} - \Sigma_q(p)
 \, , \qquad (q=u,d,s) \, ,
\label{gapDSE}
\end{equation}
(while $S^{\scriptstyle {\rm free}}_{q}$ are free ones),  
and the Bethe-Salpeter equation (BSE) for the $q{\bar q}'$ meson bound-state vertices 
$\Gamma_{q{\bar q}'}$:
\begin{eqnarray}
\label{BSE}
\Gamma_{q{\bar q}'}(k,{p})_{ef} =  \qquad \qquad \qquad \qquad
\qquad \qquad \qquad \quad \quad
\\  
\int \!
[S_q(\ell+\frac{{p}}{2}) \Gamma_{q{\bar q}'}(\ell,{p})
S_{q'}(\ell-\frac{{p}}{2}) ]_{gh} K(k-\ell)_{ef}^{hg}
\frac{d^4\ell}{(2\pi)^4}~,   \nonumber
\end{eqnarray}
where $K$ is the interaction kernel, and $e,f,g,h$ represent (schematically)
collective spinor, color and flavor indices.

This nonperturbative and covariant bound-state DS approach can be applied at 
various degrees of truncations, assumptions and approximations, ranging from {\it
ab initio} QCD calculations and sophisticated truncations ({\it e.g.,} see
\cite{Alkofer:2000wg,Roberts:2000aa,Holl:2006ni,Fischer:2006ub,Eichmann:2014xya,Binosi:2016rxz,Qin:2016fwx} and references therein)
to very simplified modeling of hadron phenomenology, such as utilizing
Nambu--Jona-Lasinio point interaction. 
For applications in involved contexts such as nonzero temperature or density,
strong simplifications are especially needed for tractability. This is why 
the separable approximation \cite{Blaschke:2000gd} is adopted presently
 [see more between Eqs. (\ref{DS-equation})-(\ref{M2_prop_m})].
However, for describing pseudoscalar mesons (including $\eta$ and $\eta^\prime$), 
reproducing the correct chiral behavior of QCD is much more important than 
dynamics-dependent details of their internal bound-state structure.

As a rarity among bound-state approaches,
the DS one can achieve the correct QCD chiral behavior --
also regardless of details of modeling dynamics, but under the condition of a
consistent truncation of DS equations, respecting pertinent Ward-Takahashi
 identities \cite{Alkofer:2000wg,Roberts:2000aa,Holl:2006ni,Fischer:2006ub}. 
A consistent DS truncation, where DChSB is very well understood, is the 
rainbow-ladder approximation (RLA). Since it also enables tractable calculations, 
it is still the most usual approximation in phenomenological applications,
and we also adopt it here. 
In RLA, the BSE (\ref{BSE}) employs the dressed quark propagator solution 
$S(p)$ from the gap equation (\ref{gapDSE})\&(\ref{DS-equation}), which 
in turn employs the same effective interaction kernel as the BSE. 
It has the simple gluon-exchange form, where both quark-gluon vertices are bare:
\begin{equation}
[K(k)]_{ef}^{hg} = {\rm i} \,
      g^2 D_{\mu\nu}^{ab}(k)_{\mbox{\rm\scriptsize eff}} \,
[\frac{\lambda^a}{2}\,\gamma^{\mu}]_{eg} \,
[\frac{\lambda^b}{2}\,\gamma^{\nu}]_{hf} \, ,
\label{RLAkernel}
\end{equation}
so that the quark self-energy in the gap equation is
\begin{eqnarray}
 \label{DS-equation}
\!\!\!\! \Sigma_q(p) =  
- \int \!\!\frac{d^4\ell}{(2\pi)^4} \,
  g^2 D_{\mu\nu}^{ab}(p-\ell)_{\mbox{\rm\scriptsize eff}} \, 
\frac{\lambda^a}{2}\,\gamma^{\mu}
S_q(\ell) \frac{\lambda^b}{2}\,\gamma^{\nu}  , 
\end{eqnarray}
where $D_{\mu\nu}^{ab}(k)_{\mbox{\rm\scriptsize eff}}$ is an 
effective gluon propagator.

These simplifications should be compensated by modeling the effective gluon 
propagator $D_{\mu\nu}^{ab}(k)_{\mbox{\rm\scriptsize eff}}$ in order to 
reproduce well the relevant phenomenology; here,  
pseudoscalar ($P$) meson  masses $M_P$,
 decay constants $f_P$, and condensates $\langle{\bar q}q\rangle$,
 including $T$-dependence of all these.
In the present paper, we use the same model as in Ref. \cite{Benic:2011fv},
whose approach to the $T$-dependence of $U_A(1)$ anomaly we now seek to improve.
All details on the functional form and parameters of this model interaction
can be found in the subsection II.A of Ref. \cite{Horvatic:2007qs}.  
Such models, so-called rank-2 separable, are phenomenologically successful
(see, {\it e.g.}, Refs.
\cite{Blaschke:2000gd,Horvatic:2007qs,Blaschke:2006ss,Horvatic:2007wu,Horvatic:2007mi}),
except they have the well-known drawback of predicting a somewhat too low
transition temperature: the model we use presently and in Refs.
\cite{Horvatic:2007qs,Horvatic:2007wu,Horvatic:2007mi,Benic:2011fv}, has
$T_{\rm Ch}=128$ MeV, {\it i.e.}, some 17\% below the now widely accepted central
value of $154\pm 9$ MeV \cite{Bazavov:2011nk,Dick:2015twa,Bazavov:2017dus}.
But, rather than quantitative predictions at specific absolute temperatures,
we are interested in the relative connection between the chiral restoration
temperature $T_{\rm Ch}$ and the temperature scales characterizing signs of
effective disappearance of the $U_A(1)$ anomaly, for which the present model
is adequate. In addition, Ref. \cite{Horvatic:2010md} shows that coupling
to the Polyakov loop can increase $T_{\rm Ch}$, while qualitative
features of the $T$-dependence of the model are preserved. Thus,
separable model results at $T > 0$ are most meaningfully presented
as functions of the relative temperature $T/T_{\rm Ch}$,
as in Refs. \cite{Horvatic:2007qs,Benic:2011fv}.

\begin{figure}[b]
\centering
\includegraphics[scale=0.30]{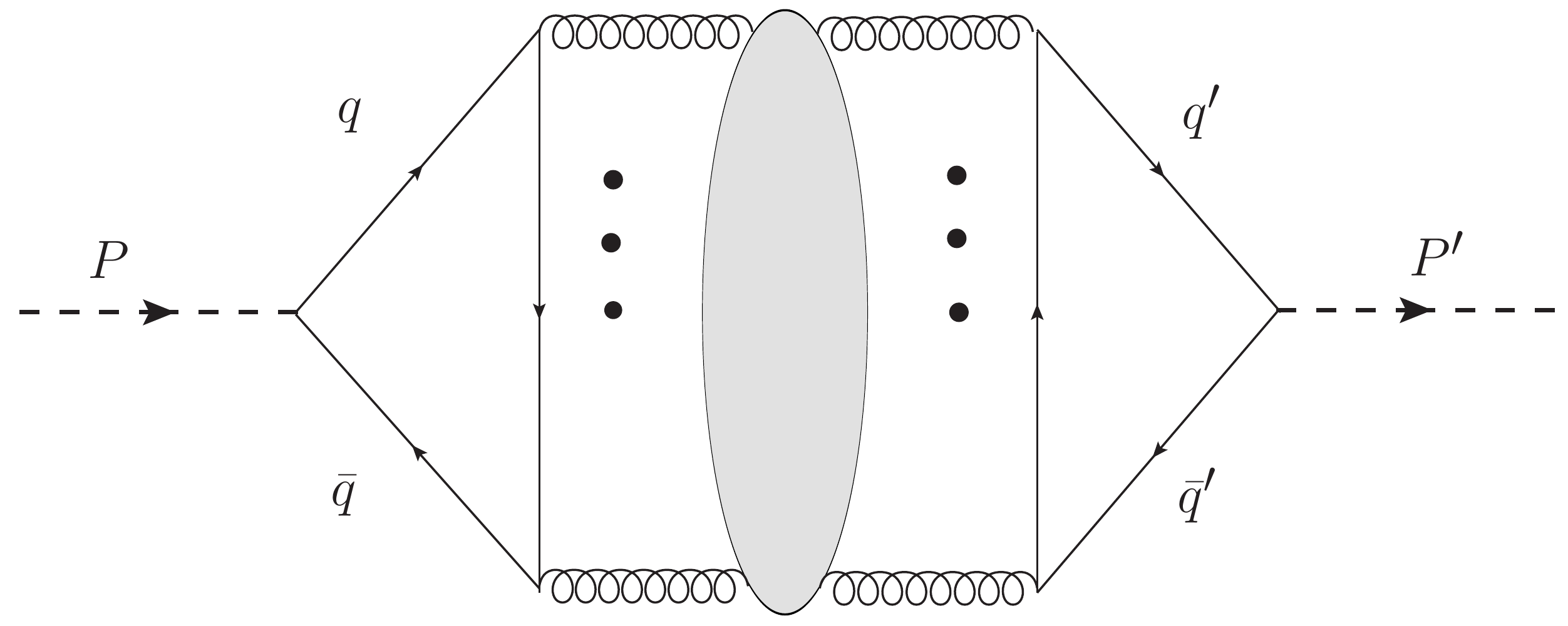}
\caption{Axial-anomaly-induced, flavor-mixing transitions from hidden-flavor 
pseudoscalar states $P=q\bar q$ to $P'=q'\bar q'$ include both possibilities 
$q=q'$ and $q\neq q'$. All lines and vertices are dressed. The gray blob 
symbolizes all possible intermediate gluon states enabling this.
 Three bold dots symbolize an even \cite{Kekez:2000aw}, 
but otherwise unlimited number of additional gluons. As pointed out in Ref.
\cite{Kekez:2000aw}, the diamond graph
 is just the simplest example of such a contribution.}
\label{Blob}
\end{figure}

Anyway, regardless of details of model dynamics, {\it i.e.,} of a choice
of $D_{\mu\nu}^{ab}(k)_{\mbox{\rm\scriptsize eff}}$, but just thanks to
the consistent truncation of DS equations, the BSE (\ref{BSE}) yields
the masses $M_{q\bar q'}$ of pseudoscalar $P\sim q{\bar q}'$ mesons which
satisfy the Gell-Mann-Oakes-Renner--type relation with the current masses
 ${m}_q$, ${m}_{q'}$ of the corresponding quarks:
\begin{equation}
M_{q\bar q'}^2 = {\rm const} \, ({m}_q + {m}_{q'}) \, ,
\qquad (q,q'=u,d,s) \, .
\label{M2_prop_m}
\end{equation}
While this guarantees all $M_{q\bar q'} \to 0$ in the chiral limit, it
also shows that RLA cannot lead to any $U_A(1)$-anomalous contribution
responsible for $\Delta M_{\eta_0}$. 
That is, RLA gives us only the non-anomalous part ${\hat M}^2_{N\!A}$ of 
the squared-mass matrix ${\hat M}^2={\hat M}^2_{N\!A}+{\hat M}^2_{A}$
of the hidden-flavor ($q=q'$) light ($q=u,d,s$) pseudoscalar mesons. In 
the basis $\{u\bar{u},d\bar{d},s\bar{s}\}$, ${\hat M}^2_{N\!A}$ is simply 
${\hat M}^2_{N\!A}={\rm diag}[M_{u\bar{u}}^2,M_{d\bar{d}}^2,M_{s\bar{s}}^2]$.
 The anomalous part ${\hat M}^2_{A}$ arises because the pseudoscalar
hidden-flavor states $q\bar{q}$ are not protected from the flavor-mixing
QCD transitions (through anomaly-dominated pseudoscalar gluonic intermediate
states), depicted in Fig. \ref{Blob}. They are obviously beyond the reach
of RLA and horrendously hard to calculate. Nevertheless, they 
cannot be neglected, as can be seen in 
the Witten-Veneziano relation (WVR) \cite{Witten:1979vv,Veneziano:1979ec}
which remarkably relates the full-QCD quantities ($\eta'$, $\eta$ and 
$K$-meson masses $M_{\eta',\eta,K}$ and the pion decay constant $f_\pi$), to
the topological susceptibility $\chi_{\rm Y\!M}$ of the (pure-gauge) YM theory:
\begin{equation}
M_{\eta'}^2 + M_\eta^2 - 2 M_K^2 = 2 N_f \,\, \frac{\chi_{\rm Y\!M}}{f_\pi^2} 
 = M_{U_A(1)}^2  \, .
\label{WittenVenez}
\end{equation}
Namely, its chiral-limit-nonvanishing right-hand-side (RHS) is large, 
roughly 0.8 to 0.9 GeV$^2$, while Eq. (\ref{M2_prop_m}) leads basically 
to the cancellation of all chiral-limit-vanishing contributions on the
 left-hand-side (LHS) \cite{Benic:2011fv}. RHS is the WVR result for
 the total mass contribution of the $U_A(1)$ anomaly to the
 $\eta$-$\eta'$ complex, $M_{U_A(1)}$.

The ${\hat M}^2_{A}$ matrix elements generated by the
$U_A(1)$-anomaly-dominated transitions $q\bar q \to q' \bar q'$
(see Fig. \ref{Blob}) can be written \cite{Kekez:2005ie}
in the flavor basis $\{u\bar{u},d\bar{d},s\bar{s}\}$ as
\begin{equation}
\langle q\bar q | {\hat M}^2_A |q' \bar q' \rangle =
b_q \, b_{q'} \,\, , \qquad (q, q' = u, d, s) \,\, .
\label{elementM2AqqX}
\end{equation}
Here $b_q = \sqrt{\beta}$ for both $q = u, d$ since we assume 
${m}_u = {m}_{d} \equiv {m}_{l}$, {\it i.e.,} isospin $SU(2)$ symmetry, 
which is an excellent approximation for most purposes in hadronic physics.
 For example,
$M_{u\bar{u}} = M_{d\bar{d}} \equiv M_{l\bar{l}} = M_{u\bar d}\equiv M_{\pi}$
obtained from BSE (\ref{BSE}) is our RLA model pion mass for
$\pi^+ (\pi^-) = u\bar d (d\bar u)$ and $\pi^0=(\, u\bar{u} - d\bar{d}\, )/\sqrt{2}$,
so that ${\hat M}^2_{N\!A}={\rm diag}[M_{\pi}^2,M_{\pi}^2,M_{s\bar{s}}^2]$.
It still contains $M_{s\bar{s}}$, the mass of the unphysical, but theoretically
very useful $s{\bar s}$ pseudoscalar obtained in RLA. However, thanks to
Eq. (\ref{M2_prop_m}), it can also be expressed through the masses of physical
mesons, $M_{s\bar s}^2 = 2 M_{u\bar s}^2 - M_{u\bar d}^2 = 2 M_K^2 - M_{\pi}^2$,
in a very good approximation
\cite{Klabucar:1997zi,Kekez:2000aw,Kekez:2001ph,Kekez:2005ie,Horvatic:2007qs,Horvatic:2007mi}.
Its decay constant $f_{s\bar s}$ is calculated in the same way as $f_\pi$ and $f_K$.

Since the $s$-quark is much heavier than $u$ and $d$ ones, in Eq.
(\ref{elementM2AqqX}) we have $b_q = X \sqrt{\beta}$ for $q = s$, with $X<1$.
Namely, transitions to and from more massive $s$-quarks are suppressed, and the
quantity $X$ expresses this influence of the $SU(3)$ flavor symmetry breaking.
The most usual choice for the flavor-breaking parameter had been
\cite{Horvatic:2007qs,Benic:2011fv,Klabucar:1997zi,Kekez:2000aw,Kekez:2001ph,Kekez:2005ie,Horvatic:2007mi}
the educated estimate $X=f_\pi/f_{s\bar s}$, but we found \cite{Benic:2014mha}
it necessarily follows in the variant of our approach relying on Shore's
\cite{Shore:2006mm,Shore:2x} generalization of WVR (\ref{WittenVenez})
-- see Sec. \ref{extensTge0}.

The anomalous mass matrix ${\hat M}^2_A$, which is of the pairing form 
(\ref{elementM2AqqX}) in the hidden-flavor basis $\{u\bar{u},d\bar{d},s\bar{s}\}$,
in the octet-singlet basis $\{\pi^0,\eta_8,\eta_0 \}$ of hidden-flavor
pseudoscalars becomes
\begin{equation}
\! {\hat M}^2_A = 
\beta \left[ \begin{array}{ccl} 0 & 0 & \qquad \quad 0 \\
\\
      0 & \frac{2}{3}(1-X)^2 & \frac{\sqrt{2}}{3}(2-X-X^2) \\
\\  
      0 & \frac{\sqrt{2}}{3}(2-X-X^2) & \quad \frac{1}{3}(2+X)^2
        \end{array} \! \right],
\label{M2A80X}
\end{equation}
showing that the $SU(3)$ flavor breaking, $X\neq 1$, is necessary for the anomalous
contribution to the $\eta_8$ mass squared, $\Delta M_{\eta_8}^2 = \beta (2/3)(1-X)^2$.
In the flavor $SU(3)$ symmetric case, $X=1$, only the $\eta_0$ mass receives
a $U_A(1)$ anomaly contribution: $M_{U_A(1)}^2 = \Delta M_{\eta_0}^2 = 3\beta$ in
this limit. Otherwise, $M_{U_A(1)}^2\equiv\text{Tr}\,{\hat M}^2_A=(2+X^2)\beta$.

The $SU(3)$ breaking, $X\neq 1$, causes ${\hat M}^2_A$ (\ref{M2A80X})
 to be off-diagonal,
but in this basis also the $\{\eta_8,\eta_0 \}$-submatrix of ${\hat M}^2_{NA}$
gets strong, negative off-diagonal elements,
$M_{80}^2 = \sqrt{2} \, ( M_{\pi}^2 - M_{s\bar{s}}^2 )/3$ (see, {\it e.g.},
 \cite{Kekez:2005ie}).  Eq. (\ref{M2A80X})
thus shows that the interplay of the flavor symmetry breaking ($X<1$) with anomaly 
is necessary for partial cancellation of the off-diagonal (8,0) elements in the 
complete mass matrix ${\hat M}^2={\hat M}^2_{N\!A}+{\hat M}^2_{A}$, {\it i.e.,}
for getting the physical isoscalars in a rough approximation as
 $\eta \approx \eta_8$ and $\eta' \approx \eta_0$.
How this changes with diminishing $U_A(1)$-anomaly contributions
 is exhibited in Secs. \ref{resultsTGge0} and \ref{summary}.

Since the isospin-limit $\pi^0$ decouples from the anomaly and  mixing,
only the isoscalar-subspace
2$\times$2 mass matrix ${\hat M}^2$ needs to be considered.  Even though 
${\hat M}^2$ is strongly off-diagonal in the isoscalar $\text{NS}$-$\text S$
basis $\{\eta_\text{NS}, \eta_\text{S} \}$,
\begin{equation}
\label{massMatrix}
\left[\!\! \begin{array}{c}\eta_\text{NS} \\
          \\  \eta_\text{S} \end{array} \!\! \right]
\! \equiv \! \left[
 \begin{array}{c} \! \frac{1}{\sqrt{2}}\, ( u\bar{u} + d\bar{d}\, )\\
                     \\    
                       s\bar{s} \end{array} \right]
\! \equiv \! \left[
 \begin{array}{ccc}\! \frac{1}{\sqrt{3}} &  & \sqrt{\frac{2}{3}} \\
                  \!   - \sqrt{\frac{2}{3}} &  &   \frac{1}{\sqrt{3}} 
        \end{array}
                    \right]
\! \left[\!\! \begin{array}{c}\eta_8 \\ 
   \\ \eta_0 \end{array} \!\! \right],
\end{equation}
in this basis it has the simplest form:
\begin{equation}
\label{massMatrix}
 {\hat M}^2  \equiv  \left[\!\!
        \begin{array}{cc} M_{\text{NS}}^2 & \! M_{\text{NS\,S}}^2 \\
\\
                        M_{\text{S\,NS}}^2 & \! M_{\text{S}}^2 
        \end{array}
                    \!\! \right]
    = \left[\!\!
        \begin{array}{cc} M_\pi^2+2\beta & \! \sqrt{2}\beta X \\
\\
                        \sqrt{2}\beta X  & \! M_{s\bar{s}}^2+\beta X^2 
        \end{array}
                     \!\! \right],
\end{equation}
which also shows that when the $U_A(1)$-anomaly contributions vanish
({\it i.e.,} $\beta\!\to\! 0$), the NS-S scenario is to be realized.
This means not only that the physical isoscalars end up as
$\eta \to \eta_\text{NS}$ and $\eta' \to \eta_\text{S}$, but 
that their respective masses become $M_\pi$ and $M_{s\bar{s}}$.

Our experience with various dynamical models (at $T=0$) shows
\cite{Klabucar:1997zi,Kekez:2000aw,Kekez:2001ph,Kekez:2005ie,Horvatic:2007mi}
that after pions and kaons are correctly described, a good determination of the
anomalous mass shift parameter is sufficient for Eq. (\ref{massMatrix}) to give
good $\eta'$ and $\eta$ masses, since $M_{s\bar s}^2 = 2 M_K^2 - M_{\pi}^2$
holds well.

Nevertheless, calculating the anomalous contributions ($\propto\!\beta$)
in DS approaches is a very difficult task. 
Ref. \cite{Bhagwat:2007ha} explored it by taking the calculation
beyond RLA, but had to adopt extremely schematic 
model interactions (proportional to $\delta$-functions in momenta) for both
the ladder-truncation part (\ref{RLAkernel}) and the anomaly-producing part.
Another approach \cite{Alkofer:2008et} obtained a qualitative agreement
with lattice on $\chi_{\rm Y\!M}$ (and consequently, acceptable masses of
$\eta'$ and $\eta$) by assuming that contributions to Fig. \ref{Blob} are
dominated by the simplest one, the diamond graph, if it is appropriately
dressed -- in particular, by an appropriately singular quark-gluon vertex.

We, however, take a different route, since our goal is {\it $\,$not} to figure
out on a microscopic level how breaking of $U_A(1)$ comes about, but to
phenomenologically model and study the high-$T$ behavior of masses of 
the realistic $\eta'$ and $\eta$, along with other light pseudoscalar mesons.
In DS context, the most suitable approach is then the one developed in
Refs.  \cite{Klabucar:1997zi,Kekez:2000aw,Kekez:2001ph,Kekez:2005ie,Horvatic:2007mi}
and extended to $T>0$ in Refs. \cite{Horvatic:2007qs,Benic:2011fv}.

The key is that $U_A(1)$ anomaly is suppressed in the limit of large number 
of QCD colors $N_c$ \cite{Witten:1979vv,Veneziano:1979ec}. So, in the sense of 
$1/N_c$ expansion, it is a controlled approximation to view the anomaly 
contribution as a perturbation with respect to the (non-suppressed) results 
obtained through RLA (\ref{RLAkernel})-(\ref{DS-equation}). While considering
meson masses, it is thus not necessary to look for anomaly-induced corrections
to RLA Bethe-Salpeter wavefunctions,{\footnote{It is instructive to 
recall \cite{Klabucar:1997zi,Gilman:1987ax} that nonet symmetry or broken version
thereof is in fact assumed, explicitly or implicitly, by all approaches using
the simple hidden-flavor basis $q\bar q$, {\it e.g.,} to construct the $SU(3)$
states pseudoscalar meson states $\eta_0$ and $\eta_8$ without distinguishing
 between the 
$q\bar q$ states belonging to the singlet from those belonging to the octet.
An independent {\it a posteriori} support for our approach is also that
 $\eta$ and $\eta' \to \gamma\gamma^{(*)}$ processes are described well
 \cite{Klabucar:1997zi,Kekez:2000aw,Kekez:2001ph,Kekez:2005ie}.}}
which are consistent with DChSB and with the chiral QCD-behavior (\ref{M2_prop_m})
 essential for description of pions and kaons.
The breaking of nonet symmetry by $U_A(1)$ anomaly can be introduced just on 
the level of the masses in the $\eta'$-$\eta$ complex, by adding to the 
RLA-calculated ${\hat M}^2_{N\!A}$ the anomalous contribution ${\hat M}^2_A$.
Its anomaly mass parameter $\beta$ can be obtained by fitting 
\cite{Kekez:2000aw} the empirical masses of $\eta$ and $\eta'$, or better 
-- because then no new fitting parameters are introduced -- from 
lattice results on YM topological susceptibility $\chi_{\rm Y\!M}$. 
Employing WVR (\ref{WittenVenez}) yields \cite{Kekez:2005ie,Benic:2011fv}
$\beta = \beta_{\text{WV}}$, while Shore's generalization gives (see
Sec. \ref{extensTge0}) $\beta = \beta_{\it{Sh\!o}}$ \cite{Benic:2014mha}:
\begin{equation}
\beta_{\text{WV}} \, = \, \frac{6 \, \chi_\text{YM}}{(2+X^2) \, f_\pi^2} 
\,\,\, , \, \quad
\label{betasWV+S}
\beta_{\it{Sh\!o}} = \frac{2\, A}{f_\pi^2}\approx \frac{2\, \chi_\text{YM}}{f_\pi^2}
\,\, , \,\,
\end{equation}
where $A$ is the QCD topological charge parameter, given below by 
Eq. (\ref{defA}) in terms of $q\bar q$ condensates of massive quarks,
which turns out to be crucial for a realistic $T$-dependence of 
the masses in the $\eta'$-$\eta$ complex.

\section{Extension to $T \ge 0$ }
\label{extensTge0}

Extending our treatment 
\cite{Klabucar:1997zi,Kekez:2000aw,Kekez:2001ph,Kekez:2005ie,Horvatic:2007mi}
of the $\eta'$-$\eta$ complex to $T>0$ is clearly more complicated. 
Since to the best of our knowledge there is no systematic derivation of the
$T>0$ version of either WVR (\ref{WittenVenez}) or its generalization by
Shore \cite{Shore:2006mm,Shore:2x}, it is tempting to try replacing
straightforwardly all quantities
 by their $T$-dependent versions.  In WVR, these are the 
full-QCD quantities $M_{\eta'}(T)$, $M_{\eta}(T)$, $M_K(T)$ and $f_\pi(T)$, 
but also $\chi_{\rm Y\!M}(T)$, which is a pure-gauge, YM quantity and thus
much more resistant to high temperatures than QCD quantities containing also
quark degrees of freedom. Indeed, lattice calculations indicate that the 
fall of $\chi_{\rm Y\!M}(T)$, from which one would expect the fall of the 
anomalous $\eta'$ mass, starts only at $T$ some 100 MeV (or even more) above
the (pseudo)critical temperature $T_{\rm Ch}$ for the chiral symmetry restoration
of the full QCD,              
around where decay constants
already fall appreciably. It was then shown \cite{Horvatic:2007qs} that the
straightforward extension of the $T$-dependence of the YM susceptibility 
would predict even an increase of the $\eta'$ mass around and beyond 
$T_{\rm Ch}$, contrary to experiment \cite{CsorgoVertesiSziklaiVargyas}.

\begin{figure}[t]
\centering
\includegraphics[scale=0.35]{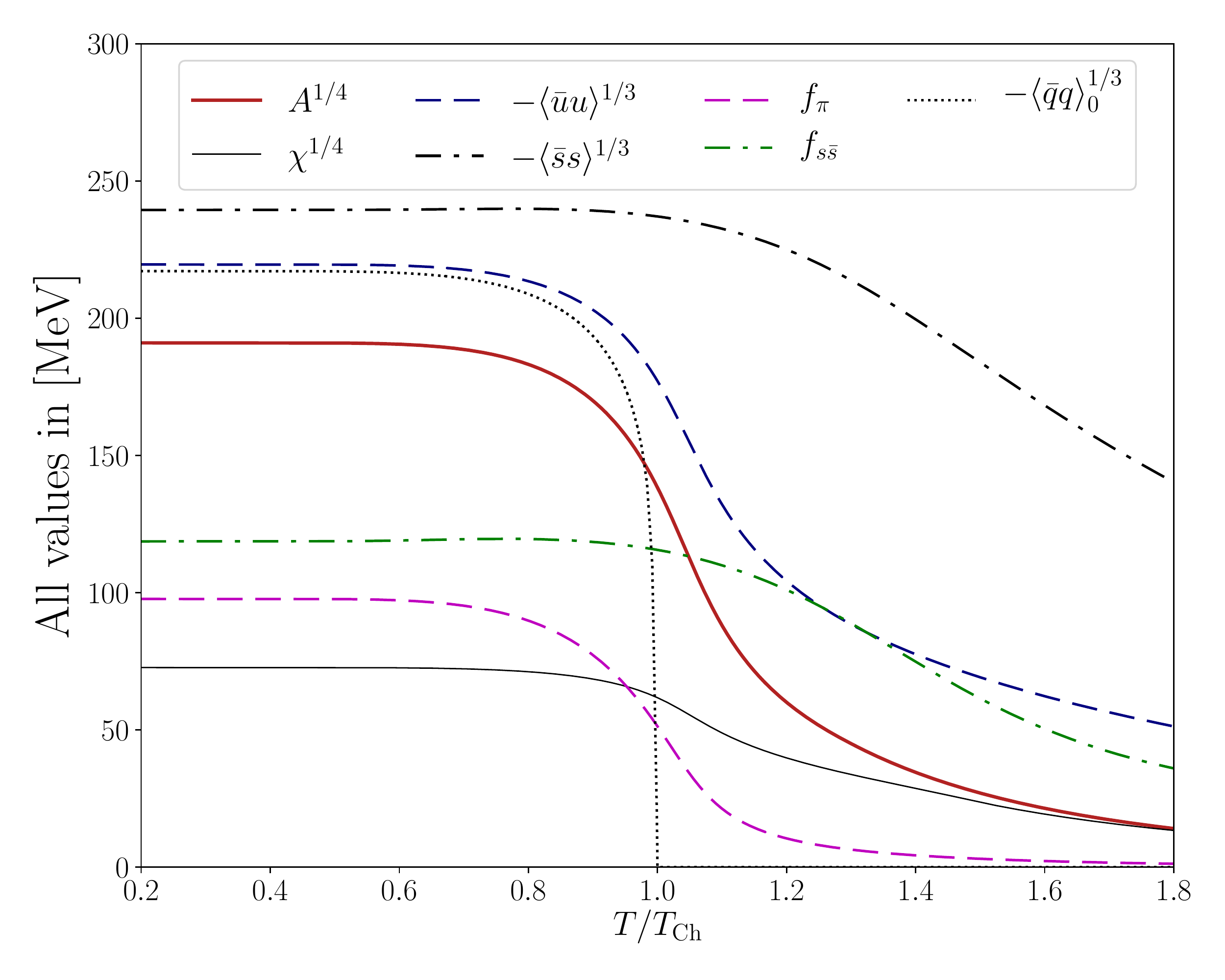}
\caption{The relative-temperature $T/T_{\rm Ch}$-dependences of the pertinent
order parameters calculated in our usual \cite{Horvatic:2007qs,Benic:2011fv}
separable interaction model.  The odd man out is
(3$^{\rm rd}$ root of the absolute value of)
 the chiral condensate $\langle{\bar q}q\rangle_0(T)$ falling steeply
at $T=T_{\rm Ch}$ and dictating similar behavior \cite{Benic:2011fv} to
${\widetilde\chi}(T)$ (\ref{LS}).
All other displayed quantities exhibit smooth, crossover behaviors, the smoother
the heavier the involved flavor is: the highest curve (dash-dotted) and the
second one from above (dashed) are (3$^{\rm rd}$ roots of the absolute values of)
the condensates $\langle{\bar s}s\rangle(T)$ and $\langle{\bar u}u\rangle(T)$,
respectively, and the resulting topological susceptibility $\chi(T)^{1/4}$
(the thin solid curve, starting as the lowest) and topological charge parameter
$A(T)^{1/4}$ (the upper, thick solid curve).
The decay constants $f_\pi(T)$ and $f_{s\bar s}(T)$ are, respectively, the
lower dashed and dash-dotted curves. (Colors online.)}
\label{condens0uds}
\end{figure}

It could be expected that at high $T$, original WVR (\ref{WittenVenez})
will not work since it relates the full-QCD quantities with a much more 
temperature-resistant YM quantity, $\chi_{\rm Y\!M}(T)$. However, this 
problem can be eliminated \cite{Benic:2011fv} by using, at $T=0$, the 
 (inverted) Leutwyler-Smilga (LS) relation \cite{Leutwyler:1992yt}:
\begin{equation}
\label{LS}
 \chi_{\rm Y\!M} \, = \, 
\frac{\chi}{1 + {\chi} \, (\,\frac{1}{ m_u} + \frac{1}{m_d} + \frac{1}{m_s} \,)
\, \frac{1}{\langle{\bar q}q\rangle_0} } \, \,
\, \equiv {\widetilde \chi}
\end{equation}
to express $\chi_{\rm Y\!M}$ in WVR (\ref{WittenVenez}) through the 
full-QCD topological susceptibility $\chi$ and the chiral-limit condensate 
$\langle{\bar q}q\rangle_0$. The zero-temperature WVR is so retained,
while the full-QCD quantities in ${\widetilde \chi}$ do not have the 
$T$-dependence mismatch with the rest of Eq. (\ref{WittenVenez}). Thus,
instead of $\chi_{\rm Y\!M}(T)$, Ref. \cite{Benic:2011fv} used at $T>0$ the
combination ${\widetilde \chi}(T)$ (\ref{LS}), where the QCD topological
susceptibility $\chi$ in the light-quark sector can be expressed
 as \cite{Di Vecchia:1980ve,Leutwyler:1992yt,Durr:2001ty}:
\begin{equation}
\chi = \frac{- \, 1}{\, (\,\frac{1}{m_u}+\frac{1}{m_d}+\frac{1}{m_s}\,)
 \,\, \frac{1}{\langle{\bar q}q\rangle_0}} \, + \, {\cal C}_m \, .
\label{chi_small_m}
\end{equation}
This implies that the (partial) restoration of $U_A(1)$ symmetry is strongly
tied to the chiral symmetry restoration, since not $\chi_{\rm Y\!M}(T)$, but
$\langle{\bar q}q\rangle_0(T)$, through ${\widetilde \chi}(T)$ (\ref{LS}),
determines the $T$-dependence of the anomalous parts of the masses in the
$\eta$-$\eta'$ complex \cite{Benic:2011fv}. The dotted curve in Fig. \ref{condens0uds}
illustrates how $\langle{\bar q}q\rangle_0(T)$ falls steeply to zero as
 $T\to T_{\rm Ch}$, {indicative of the $2^{\rm nd}$ order phase transition}.
This behavior is followed closely by ${\widetilde \chi}(T)$, and therefore also by the
anomaly parameter $\beta_{\rm WV}(T)$ (\ref{betasWV+S}). This makes the mass matrix
(\ref{massMatrix}) diagonal immediately after $T = T_{\rm Ch}$, which marks the
abrupt onset of the NS-S scenario $M_{\eta'}(T)\to M_{s\bar s}(T)$,
$M_{\eta}(T)\to M_\pi(T)$ \cite{Benic:2011fv}.

In Eq. (\ref{chi_small_m}), ${\cal C}_m$ denotes corrections of higher orders
in small $m_q$, but should not be neglected, as ${\cal C}_m \neq 0$ is needed 
to have a finite $\chi_{\rm Y\!M}$ with Eqs. (\ref{LS})-(\ref{chi_small_m}).
They in turn give us the value ${\cal C}_m$ at $T=0$ in terms of $q\bar q$
condensate and the YM topological susceptibility $\chi_{\rm Y\!M}$.
However, to the best of our knowledge, the functional form of ${\cal C}_m$
is not known. Ref. \cite{Benic:2011fv} thus tried various parameterizations
covering reasonably possible $T$-dependences of ${\cal C}_m(T)$, but this
did not affect much the results for the $T$-dependence of the masses in 
the $\eta'$-$\eta$ complex.

An alternative to WVR (\ref{WittenVenez}) is its generalization by Shore
 \cite{Shore:2006mm,Shore:2x}. There, relations containing the masses of the
pseudoscalar nonet mesons take into account that $\eta$ and $\eta'$ should have two
decay constants each \cite{2decConst}. 
If one chooses to use the $\eta_8$-$\eta_0$ basis, they are
$f^8_{\eta},f^8_{\eta'},f^0_{\eta},f^0_{\eta'}$, and can be equivalently expressed
through purely octet and singlet decay constants ($f_8,f_0$) and two mixing angles
($\theta_8,\theta_0$). This may seem better suited for usages with effective meson
Lagrangians than with $q\bar q'$ substructure calculations starting from the
 (flavor-broken) nonet symmetry, such as ours. Nevertheless, Shore's approach was
 adapted also to the latter bound-state context, and successfully applied there --
 in particular, to our DS approach in RLA \cite{Horvatic:2007mi}. This was thanks
 to applying the simplifying scheme of Feldmann, Kroll and Stech (FKS)
 \cite{FeldmannKrollStech98PRD+FeldmannKrollStech99PLB,Feldmann:1999uf}. They
 showed that this ``2 mixing angles for 4 decay constants" formulation in the
$\text{NS-S}$ basis, although in principle equivalent to the $\eta_8$-$\eta_0$ 
basis formulation, can in practice be more simplified down to one-mixing-angle
scheme using plausible approximations based on the Okubo-Zweig-Iizuka (OZI) rule.
 Namely, the {decay-constant mixing angles} in this basis are mutually close, 
$\phi_{\text S} \approx \phi_{\rm NS}$, {\it and} both approximately equal to
the {\it state} mixing angle $\phi$ rotating the $\text{NS-S}$ basis states
 into the physical $\eta$ and $\eta'$ mesons,
\begin{equation}
\eta = \cos\phi \, \eta_{\text{NS}}
             - \sin\phi \, \eta_\text{S}~,
\,\,\,\,
\eta^\prime = \sin\phi \, \eta_\text{NS}
             + \cos\phi \, \eta_\text{S}~, 
\label{MIXphi}
\end{equation}
which diagonalizes the mass (squared) matrix (\ref{massMatrix}).

So, Ref. \cite{Horvatic:2007mi} solved numerically Shore's equations (combined
with the FKS approximation scheme) for meson masses for several dynamical DS 
bound-state models \cite{Kekez:2000aw,Kekez:2005ie,Horvatic:2007qs}. Then, Ref.
\cite{Benic:2014mha} presented analytic solutions thereof, for the masses 
of $\eta$ and $\eta'$ and the state $\text{NS-S}$ mixing angle $\phi$. These 
are longish, but closed-form expressions in terms of non-anomalous meson masses 
$M_\pi$, $M_K$ and their decay constants $f_\pi,f_K$, but also $f_\text{NS}$ 
and $f_\text{S}$, the decay constants of the unphysical $\eta_\text{NS}$ and 
$\eta_\text{S}$, and, most notably, of the full QCD topological charge parameter
$A$. This is the quantity, taken over \cite{Shore:2006mm,Shore:2x}
from Di Vecchia and Veneziano \cite{Di Vecchia:1980ve}, which in the mass relations
 of Shore's generalization has the role of $\chi_\text{YM}$ in WVR. $A$ will be
 considered in detail for the $T>0$ extension, but now let us note that
 although Shore's generalization is in principle valid to all orders in $1/N_c$
 \cite{Shore:2006mm,Shore:2x}, 
 Shore himself took advantage of
\begin{equation}
\! A = \chi_\text{YM} + {\cal O}(\frac{1}{N_c})  
 \qquad (\text{at} \,\, T=0 ) ,  
\label{AapproxChiYM}
\end{equation}
and approximated $A$, as shall we at $T=0$, by the lattice result 
$\chi_\text{YM} = (0.191 \, \rm{GeV})^4$ \cite{DelDebbio:2004ns}.

Further, one should note that since the FKS scheme neglects OZI-violating
contributions, that is, gluonium admixtures in $\eta_\text{NS}$ and
$\eta_\text{S}$, it is consistent to treat them as pure $q\bar q$ states,
accessible by our BSE (\ref{BSE}) in RLA.
Then $f_\text{NS}=f_\pi$, and $f_\text{S}=f_{s\bar s}$, the decay constant of the
aforementioned ``auxiliary" RLA $s{\bar s}$ pseudoscalar. We calculate its mass 
$M_{s\bar s}$ through BSE, but at $T=0$ it can also be related to the measurable
pion and kaon masses, $M_{s\bar s}^2 \approx 2 M_K^2 - M_{\pi}^2$, due to
Eq. (\ref{M2_prop_m}).
Similarly, $f_{s\bar s}$ can also be approximately expressed by these
 measurable quantities as $f_{s\bar s} \approx 2 f_K - f_\pi$.
Thus, up to taking $A\approx \chi_\text{YM}$ from lattice, Ref. \cite{Benic:2014mha}
could calculate the $\eta$-$\eta'$ complex using in its analytic solutions both the
 model-calculated, and {\it also} the empirical $M_\pi$, $M_K$, $f_\pi$ and $f_K$.
 So, it \cite{Benic:2014mha} checked (independently of any model) the soundness
 of our approach at $T=0$.

The analytic solutions of Ref. \cite{Benic:2014mha} also lead to the simple
elements of the mass matrix (\ref{massMatrix}):
\begin{eqnarray}
\label{NS-NSS}
M_{\text{NS}}^2 &=&  M_\pi^2 + \frac{4A}{f_\pi^2}  \,\, ,   \qquad
M_{\text{NS\,S}}^2  =  \frac{2\sqrt{2}A}{f_\pi f_{s\bar{s}}}
\\
\label{S}
M_{\text{S}}^2  &=&  M_{s\bar s}^2 + \frac{2A}{f_{s\bar{s}}^2} \,\, ,
\end{eqnarray}
implying $\, X=f_\pi/f_{s\bar{s}} \,\, $,  
$\, M_{U_A(1)}^2 = {4A}/{f_\pi^2} + {2A}/{f_{s\bar{s}}^2} \,$ 
and $\, \beta_{\it{Sh\!o}}\,$ in Eq. (\ref{betasWV+S}).
The approximation $\, A = \chi_\text{YM}\, $ (\ref{AapproxChiYM}) with
$\, \chi_\text{YM} = (0.191 \, \rm{GeV})^4\, $ from lattice \cite{DelDebbio:2004ns}
then yields 
 $\, M_{\eta'}=997 \, \rm{MeV}$ and $\, M_{\eta}=554 \, \rm{MeV}$ at $T=0$.

Since the adopted DS model enables the calculations of
non-anomalous $q\bar q$ masses and decay constants also for $T>0$, the
only thing still missing is the $T$-dependence of the full QCD topological
charge parameter $A$, as $\chi_\text{YM}(T)$ is inadequate. 
But, $A$ is used to express the QCD susceptibility $\chi$ through the 
``massive" condensates $\langle {\bar u}u \rangle$, $\langle{\bar d}d \rangle$
 and $\langle {\bar s}s \rangle$, {\it i.e.}, away from the chiral limit, 
in contrast to relations (\ref{LS}) and (\ref{chi_small_m}),
{\it e.g.}, see Eq. (2.12) in Ref. \cite{Shore:2006mm}.
Its inverse, expressing $A$, thus also contains the $q\bar q$ condensates
out of the chiral limit for all light flavors $q=u,d,s$, 
\begin{equation}
\label{defA}
A \, = \, \frac{\chi}{\,\, 1 \, +\, {\chi} \, (\,\frac{1}{m_{u}\,\langle{\bar u}u\rangle}
 + \frac{1}{m_{d} \,\langle {\bar d}d \rangle}
+ \frac{1}{m_{s}\,\langle {\bar s}s \rangle } \,) \,} \, ,
\end{equation}
and so should $\chi$ in (\ref{defA}). That is, the light-quark expression for the QCD
topological susceptibility in the context of Shore's approach should be expressed by 
the current masses $m_q$ multiplied by respective condensates $\langle {\bar q}q \rangle$
realistically away from the chiral limit:
\begin{equation}
\chi \, = \, \frac{- \, 1}{\,\,\,  \frac{1}{\, m_{u} \, \langle {\bar u}u \rangle } +
\frac{1}{m_{d}\,\langle{\bar d}d \rangle} + \frac{1}{m_{s}\,\langle {\bar s}s \rangle }
            \,\,\,  } \, + \, {\cal C}_m \, .
\label{chiShore_small_m}
\end{equation}
As before \cite{Benic:2011fv}, the small-magnitude and 
necessarily negative correction term ${\cal C}_m$ is found
 by assuming $A = \chi_\text{YM}$ at $T=0$. This large-$N_c$ approximation
 also recovers the LS relation (\ref{LS}) easily:
by approximating the realistically massive condensates with
$\langle{\bar q}q\rangle_0$ everywhere in Eq. (\ref{defA}), the QCD
topological charge parameter $A$ reduces to $\widetilde \chi$, 
justifying the conjecture of Ref. \cite{Benic:2011fv} tying the
 $U_A(1)$ symmetry restoration with the chiral symmetry one.

\begin{figure}[t]
\centering
\includegraphics[scale=0.42]{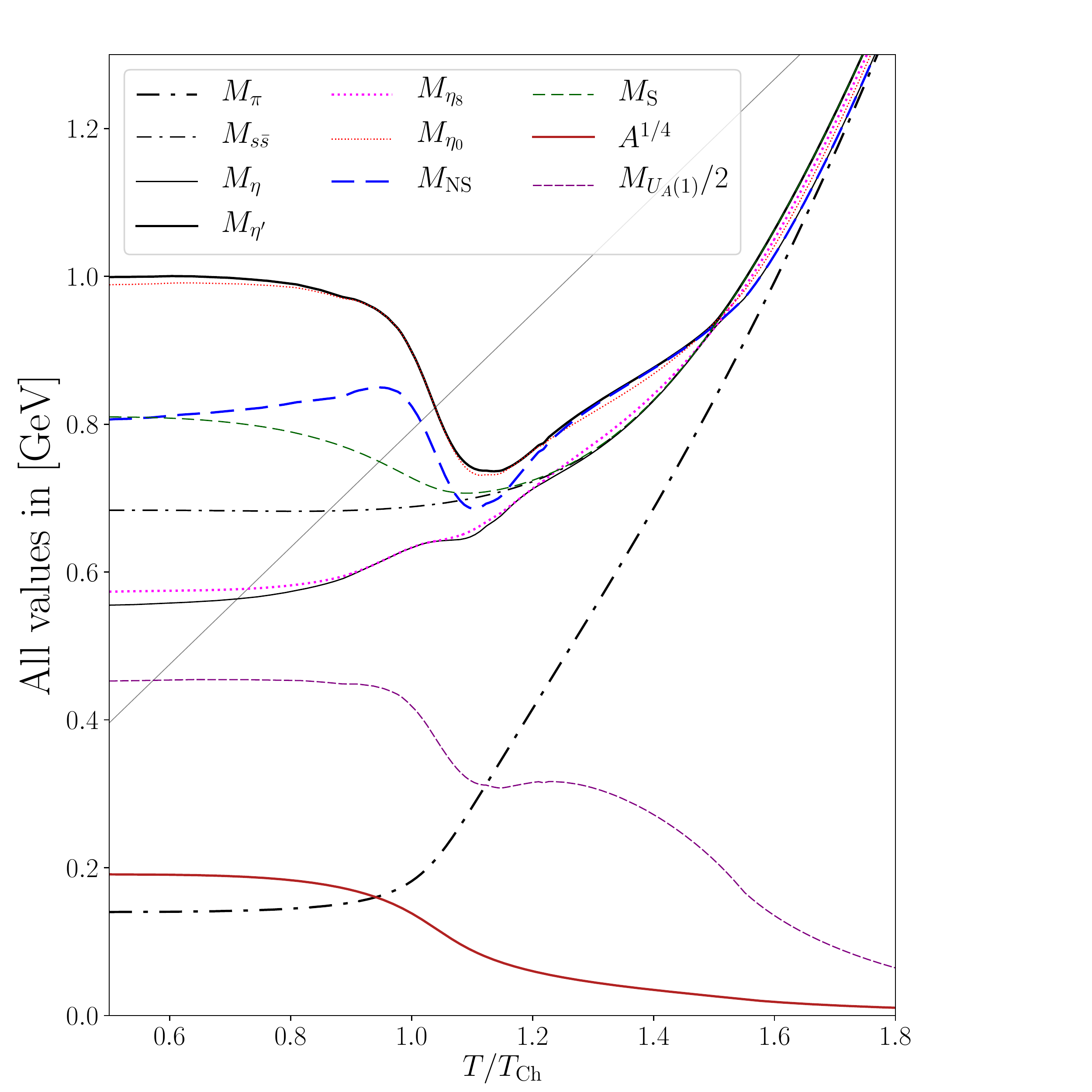}
\caption{$T$-dependence, relative to $T_{\rm Ch}$, of various $\eta'$-$\eta$ complex
 masses described in the text, $\pi$ mass (thick, lower dash-dotted curve)
 for reference,
 the {\it halved} (to avoid crowding of curves) total $U_A(1)$-anomaly-induced mass
$\frac{1}{2} M_{U_A(1)}$ (lower short-dashed curve), and topol. ch. parameter
$A^{1/4}$ as the lowest solid curve. The straight line is $2\times$ lowest
fermion Matsubara frequency $2\pi T$. (Colors online.)
}
\label{MwCT}
\end{figure}

\begin{figure}[t]
\centering
\includegraphics[scale=0.42]{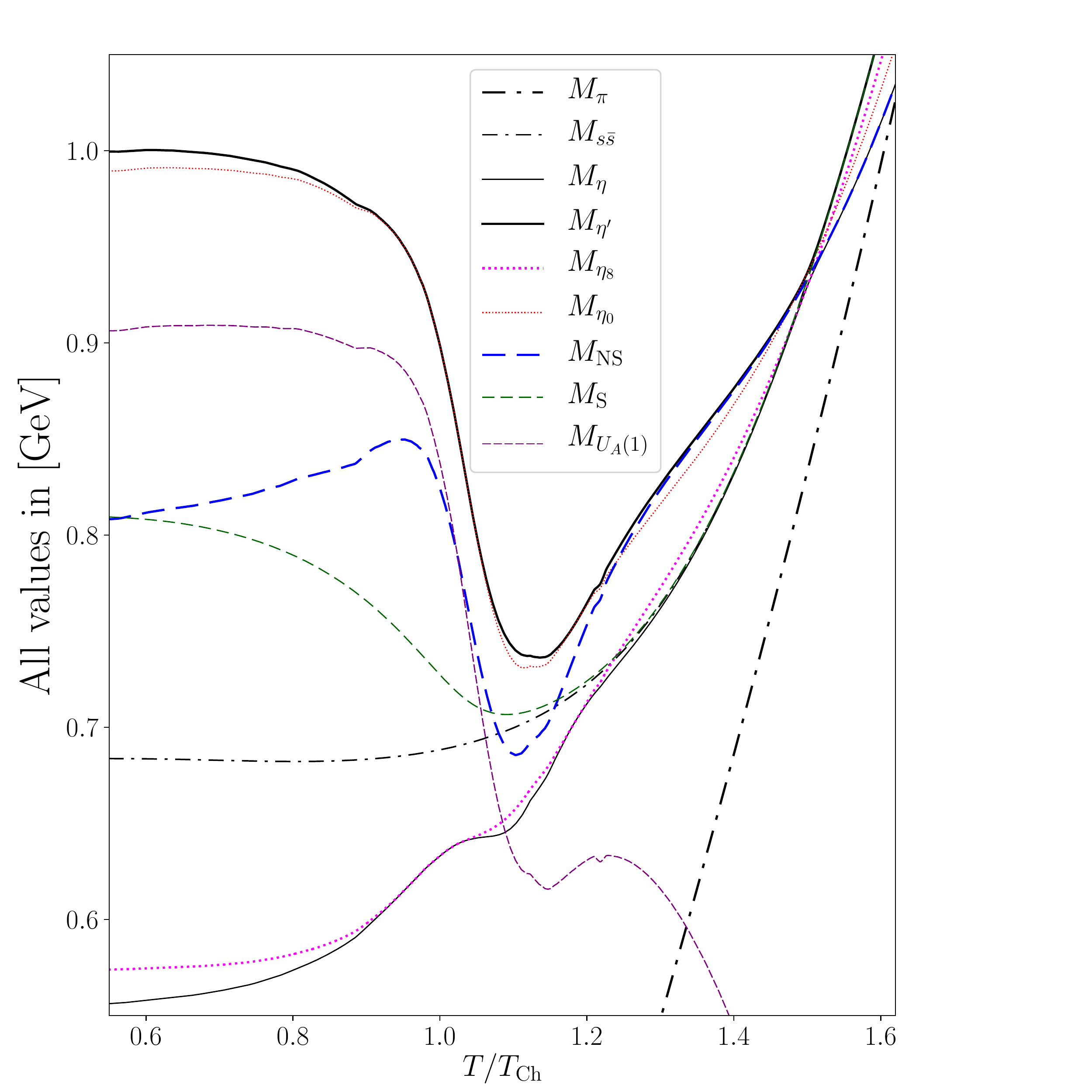}
\caption{$T/T_{\rm Ch}$-dependence of pseudoscalar meson masses zoomed to the area
important for the $\eta'$-$\eta$ complex, for the simplest {\it Ansatz}
$\, {\cal C}_m(T)= {\rm constant}={\cal C}_m(0)$, which limits temperatures to
$T \lesssim 1.6\, T_{\rm Ch}$.
}
\label{MwC0}
\end{figure}

This connection between the two symmetries is still present. However, with the
massive condensates we also get a more realistic, crossover
$T$-dependence of the masses, depicted in Figs. \ref{MwCT} and \ref{MwC0},
and presented in Sec. \ref{resultsTGge0}.

The two Figs. \ref{MwCT} and \ref{MwC0} correspond to two variations of the unknown
$T$-dependence ${\cal C}_m(T)$ of the correction term in Eq. (\ref{chiShore_small_m}).
As in Ref. \cite{Benic:2011fv}, the simplest {\it Ansatz} is constant,
 ${\cal C}_m(T)={\cal C}_m(0)$, which is most reasonable for $T < T_{\rm Ch}$,
where the condensates, and thus also the leading term in $\chi(T)$, change little.
But above some higher $T$, the negative
${\cal C}_m(0)$, although initially much smaller in magnitude than the leading term,
will make $\chi(T)$ (\ref{chiShore_small_m}), and therefore also $A(T)$, change sign.
Concretely, this limiting $T$ above which there is no meaningful description
is found a little above $1.6\, T_{\rm Ch}$.

For another, non-constant ${\cal C}_m(T)$ that would not have such a limiting
temperature, we now have a lead from lattice where the high-$T$ asymptotic behavior
 of the QCD topological susceptibility has been found to be a power law,
 $\chi(T)\propto T^{-b}$ \cite{Petreczky:2016vrs,Borsanyi:2016ksw}. The high-$T$
 dependence of our model-calculated condensates is also, without fitting,
such that the  leading term of our $\chi(T)$ in Eq. (\ref{chiShore_small_m})
 has the similar power-law behavior, with $b=5.17$.  Also, the values of our
 leading term are, qualitatively, for all $T$ roughly in the same ballpark
 as the lattice results \cite{Petreczky:2016vrs,Borsanyi:2016ksw}.
 We thus fit the quickly decreasing power-law ${\cal C}_m(T)$ for high $T$ requiring:
 {\it (i)} that this more or less rough consistency with lattice $\chi(T)$-values
 is preserved,
 {\it (ii)} that the whole $\chi(T)$ has the high-$T$ power-law dependence as the
 leading term (with $b=5.17$), and {\it (iii)} that ${\cal C}_m(T)$ joins smoothly
 with the low-$T$ value ${\cal C}_m(0)$ determined from $\chi_\text{YM}$ at $T=0$.

Our non-constant choice of ${\cal C}_m(T)$ yields the masses in Fig. \ref{MwCT}
(and $\chi(T)$ and $A(T)$ in Fig. \ref{condens0uds}), but these
 results turn out very similar to the ones with ${\cal C}_m(T)={\cal C}_m(0)$
(of course, only up to the limiting $T$ a little above $1.6 \, T_{\rm Ch}$), in
Fig. \ref{MwC0}. Thus, we present Fig. \ref{MwC0} on a different scale from 
Fig. \ref{MwCT}, {\it i.e.,} only the mass interval between 0.55 GeV and 1.05 GeV
to zoom on the $\eta$-$\eta'$ complex and discern better its various overlapping
 curves, including $M_{U_A(1)}(T)$.

The second choice of ${\cal C}_m(T)$ enables in principle the calculation of
$\chi(T)$ and $A(T)$ without any limiting $T$. 
Nevertheless, Fig. \ref{MwCT} does not reach higher than $T = 1.8 \, T_{\rm Ch}$,
because the model chosen for the RLA part of our calculations seems to become
unreliable at higher $T$'s. Namely, mass eigenvalues seem increasingly too high,
since they tend to cross the sum of lowest $q$+$\bar q$ Matsubara frequencies. 
Fortunately, by $T/T_{\rm Ch} = 1.8$, the asymptotic scenario for the anomaly has
been reached, as explained in the next section giving the detailed description of
all pertinent results at $T \ge 0$ in the next section.

\section{Results at $T \ge 0$ in detail}
\label{resultsTGge0}

Fig. \ref{condens0uds} shows how various magnitudes of current quark masses $m_q$
influence the $T$-dependence and size of $q\bar q$ condensates $\langle{\bar q}q\rangle$
and pseudoscalar decay constants $f_{q\bar q}$ calculated in our adopted
model. Defined, {\it e.g.}, in the subsection II.A of Ref. \cite{Horvatic:2007qs},
it employs the parameter values  ${m}_u = {m}_{d} \equiv {m}_{l}=5.49$ MeV and
 ${m}_s = 115$ MeV.
 
Both for condensates and decay constants, larger current quark masses lead 
to larger ``initial" ({\it i.e.}, $T=0$) magnitudes, and, what is even more
important for the present work, to smoother and slower falloffs with $T$.
 The magnitude of (the third root of) the strange quark condensate is the
  highest, dash-dotted curve in Fig. \ref{condens0uds}. 
Its $T=0$ value $|\langle{\bar s}s\rangle|^{1/3}=238.81$ MeV remains almost
unchanged till $T=T_{\rm Ch}$, and falls below 200 MeV, {\it i.e.,} by some 20\%,
only for $T\approx 1.5 \, T_{\rm Ch}$. On the other hand, the $T=0$ value of the
 isosymmetric condensates of the lightest flavors, $\langle{\bar u}u\rangle =
\langle{\bar d}d\,\rangle\equiv\langle\,{\bar l}\,l\rangle=(-218.69\,\rm MeV)^3$
is quite close to the chiral one, $\langle{\bar q}q\rangle_0 = (- 216.25\, \rm MeV)^3$,
showing how well the chiral limit works for $u$ and $d$ flavors in this respect.
Still, the small current masses of $u$ and $d$ quarks are sufficient to lead
to a very different $T$-dependence of the lightest condensates, depicted by the
dashed curve. It exhibits a typical smooth crossover behavior around 
$T=T_{\rm Ch}$, and while the falloff is much more pronounced than in the case
of $\langle{\bar s}s\rangle$, it differs qualitatively from the sharp drop to
zero exhibited by the chiral condensate (and thus also by anomaly-related quantity
 ${\widetilde\chi}(T)$ defined by LS relation (\ref{LS})). 

The isosymmetric pion decay constant $f_\pi(T) \equiv f_{l\bar l}(T)$ is the lower
dashed curve in Fig. \ref{condens0uds}, starting at $T=0$ from our model-calculated
value $f_\pi = 92$ MeV. It is quite fast-falling, in contrast to $f_{s\bar s}(T)$
(starting at $f_{s\bar s}(T=0)=119$ MeV), the decay constant of the unphysical, RLA
${\bar s}s$ pseudoscalar. It exhibits much ``slower" $T$-dependence, in
accordance with the $s$-quark condensate $\langle{\bar s}s\rangle(T)$.

The behavior of $m_l\,\langle{\bar l}\, l\rangle(T)$ largely determines that of
the full QCD topological charge parameter $A(T)$, depicted in Fig. \ref{condens0uds} 
by the thick solid curve, and in Fig. \ref{MwCT} by the lowest solid curve.
 Namely, $A$ is dominated by the lightest flavor, just like
$\chi$ and $\widetilde\chi$, as shown by their related defining expressions
 (\ref{defA})-(\ref{chiShore_small_m}) and (\ref{LS})-(\ref{chi_small_m}). 

 The smooth, monotonic fall of $A(T)$ after $T\sim 0.7\, T_{\rm Ch}$ reflects
 the degree of gradual, crossover restoration of the $U_A(1)$ symmetry with $T$.
 How this is reflected on the masses in the $\eta$-$\eta'$ complex, depends
 also on the ratios of $A(T)$ with $f_\pi^2(T)$, $f_\pi f_{s\bar s}(T)$ and
 $f_{s\bar s}^2(T)$ in Eqs.  (\ref{NS-NSS})-(\ref{S}).
$M_{\text{NS\, S}}^2\propto {A(T)/[f_\pi(T) f_{s\bar s}(T)]}$ decreases  
comparably to $A(T)^{1/2}$, and ${2A(T)/f_{s\bar s}(T)^2}$  even faster.
 Thus $M_{\text{S}}(T)$ (\ref{S}) goes monotonically into the anomaly-free
 $M_{s\bar s}(T)$ basically in the same way as in Ref. \cite{Benic:2011fv},
 except now this process is not completed at $T=T_{\rm Ch}$, but, due to the
 $A(T)$ crossover, it is drawn-out till $T\approx 1.15\, T_{\rm Ch}$.

\begin{figure}[t]
\centering
\includegraphics[scale=0.60]{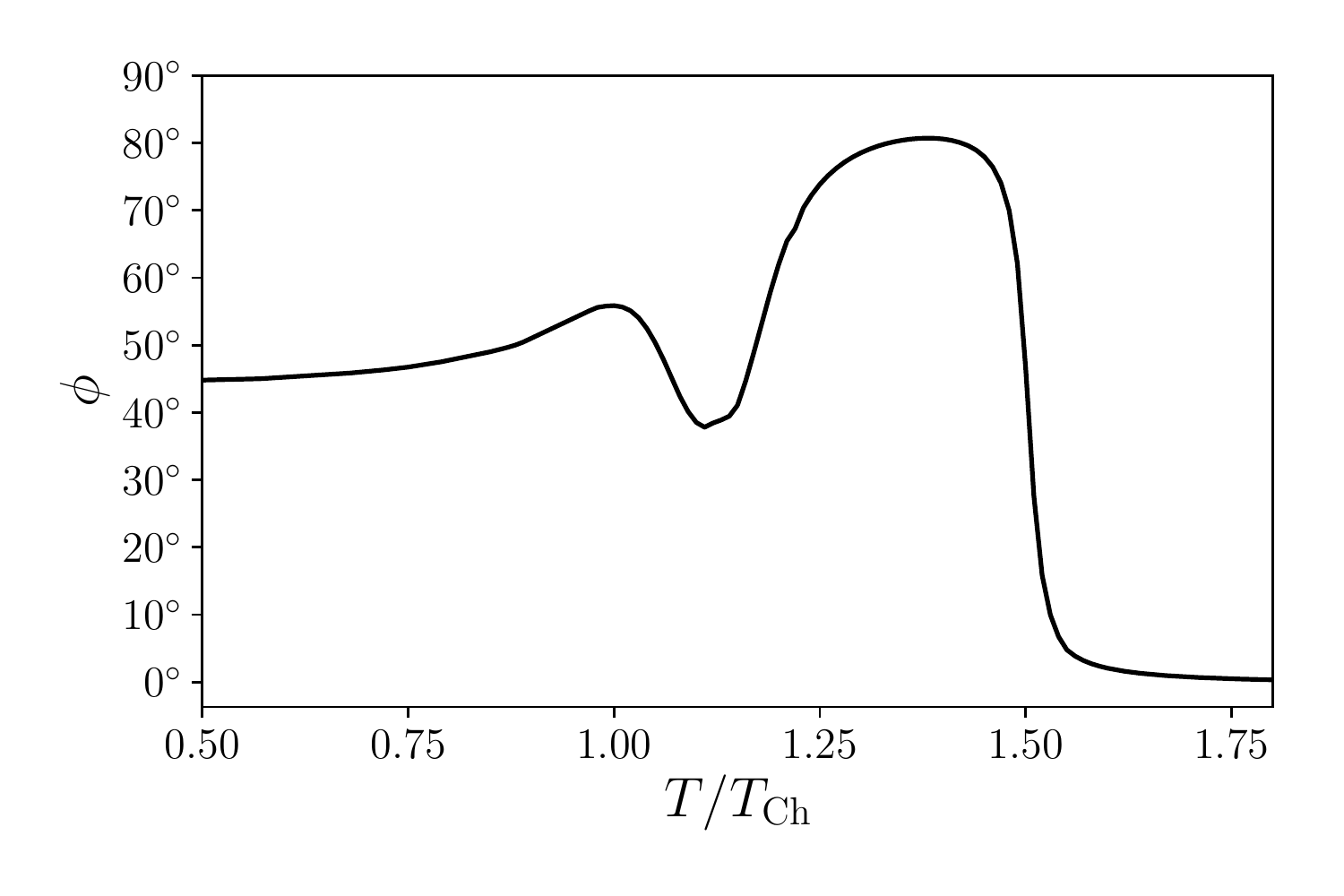}
\caption{Relative $T$-dependence of the NS-S mixing angle $\phi(T)$.  }
\label{phiTrel}
\end{figure}

In contrast, $\beta_{S\!h\!o}(T)=2A(T)/f_\pi^2(T)$ even grows for $T<0.95\, T_{\rm Ch}$
and $ 1.15\, T_{\rm Ch} \lesssim T \lesssim 1.25\, T_{\rm Ch}$.
By making $M_{\text{NS}}(T)> M_{\text{S}}(T)$ 
it causes the increase of the mixing angle $\phi$ (look at Figs. \ref{phiTrel},
 \ref{MwCT} and \ref{MwC0} together). Note that this makes the $\eta_8$-$\eta_0$
state mixing angle $\theta$ $(\approx \phi - 55^\circ)$ less negative,
{\it i.e.}, closer to zero, and brings $\eta_0$ and $\eta_8$ in an even better
 agreement with, respectively, $\eta'$ and $\eta$, than at $T=0$.

These two limited increases of $A(T)/f_\pi^2(T)$ may be model dependent 
and are not important, but what is systematic and thus important is that
the ``light" decay constant $f_\pi(T)$ is making $\sqrt{A(T)/f_\pi^2(T)}$
more resilient to $T$ not only than $A(T)^{1/4}$ itself, but also than
other anomalous mass contributions in Eqs. (\ref{NS-NSS})-(\ref{S}).

Indeed, $\beta_{S\!h\!o}(T) = 2A(T)/f_\pi^2(T)$ falls only after 
$T\approx 0.95\,  T_{\rm Ch}$ (contributing over a half of the $\eta'$ mass drop)
and then again rises somewhat after $T\approx 1.15\, T_{\rm Ch}$, to start 
definitively falling only after $T\approx 1.25\, T_{\rm Ch}$, but even then
slower than other anomalous contributions.
This makes $M_{\text{NS}}(T)$ larger enough than $M_{\text{S}}(T)$
 to rise $\phi(T)$ to around $80^\circ$, and keep it there
 as far as $T\sim 1.5\, T_{\rm Ch}$, see Fig. \ref{phiTrel}.

This explains how the masses of the physical mesons $\eta'$ and $\eta$
(thick and thin solid curves in Figs. \ref{MwCT}, \ref{MwC0}),
\begin{equation}
\! M_{\eta^\prime (\eta)}^2 = \frac{M_{\text{NS}}^2+\!M_{\text{S}}^2}{2} +(-) 
\sqrt{\left(\!\frac{M_{\text{NS}}^2-\!M_{\text{S}}^2}{2}\right)^{\!\!2}
 \!+\! M_{\text{NS\,S}}^{\,4}} \,\, ,
\label{M2etas}
\end{equation}
exhibit the mass drop of the heavier partner $\eta'$ which is almost as strong
as in the case \cite{Benic:2011fv} of the abrupt disappearance of the anomaly
contribution, while on the contrary the lighter partner $\eta$ now does not show any
sign of the mass reduction around $T = T_{\rm Ch}$, let alone an abrupt degeneracy
with the pion. The latter happens in the case with the sharp phase transition
because the fast disappearance of the {\it whole} $M_{U_A(1)}$ around $T_{\rm Ch}$
can be accommodated only by the sharp change of the state mixing ($\phi\to 0$) to
fulfill the asymptotic NS-S scenario immediately after $T_{\rm Ch}$.
(See esp. Fig. 2 in Ref. \cite{Benic:2011fv}. Note that in our approach
$M_{\eta^\prime}(T)$ cannot drop much more than a third of $M_{U_A(1)}$,
since RLA $M_{s\bar s}(T)$ is the lower limit of $M_{\eta^\prime}(T)$
both in Ref. \cite{Benic:2011fv} and here.)

In the present crossover case, however,
$T = T_{\rm Ch}$ does not mark the drastic change of the mixing of the isoscalar
states, but $\eta'$ stays mostly $\eta_0$ and $\eta$ stays mostly $\eta_8$. Then,
$\Delta M_{\eta_8}^2 = 4 A(1/f_\pi - 1/f_{s\bar s})^2/3$ [from Eq. (\ref{M2A80X})]
can serve as a compact illustration how for the lighter partner $\eta$, with $(-)$
 in Eq. (\ref{M2etas}), anomalous contributions cancel to a large extent anyway.
Thus, the mass of $\eta$ behaves mostly like the masses of other $q\bar q'$
(almost-)Goldstone bosons after losing their chiral protection at $T_{\rm Ch}$:
it just suffers the thermal rise towards $2\pi T$. 

Nevertheless, in $M_{\eta^\prime}$ (\ref{M2etas}), the anomalous contributions from
Eqs. (\ref{NS-NSS})-(\ref{S}) all add. The {\it partial} restoration of ${U_A(1)}$
symmetry around $T_{\rm Ch}$, where around a third of the total $U_A(1)$-anomalous
mass $M_{U_A(1)}$ goes away, is consumed almost entirely by the drop of
the $\eta'$ mass over the crossover.  

After $T\approx 1.15\, T_{\rm Ch}$, $M_{\eta'}(T)$ starts rising again, but
this is expected since after $T\approx T_{\rm Ch}$ light pseudoscalar mesons
start their thermal rise
towards $2\pi T$, twice the lowest Matsubara frequency of the free quark and
antiquark. This rather steep joint rise brings all the mass curves $M_P(T)$
quite close after $T\sim 1.5\, T_{\rm Ch}$. The kaon mass $M_K(T)$ is not shown
in Figs. \ref{MwCT} and \ref{MwC0} to avoid crowding of curves, but at this
temperature of the characteristic $\eta$-$\eta'$ anticrossing, $M_K(T)$ is
roughly in between $M_\pi(T)$ and the $\eta$ mass, only to be soon crossed by
$M_\eta(T)$ tending to become degenerate with $M_\pi(T)$ as detailed in the
 following passage.

   The rest of $M_{U_A(1)}(T)$, melting as $2\sqrt{A(T)}/f_\pi(T)$,
is under $1.5\, T_{\rm Ch}$ sufficiently large to keep 
$M_{\text{NS}}(T) > M_{\text{S}}(T)$ and $\phi \approx 80^\circ$.
So large $\phi$ makes $\theta$ positive, but not very far from zero,
so that still $\eta'\approx\eta_0$ and $\eta\approx\eta_8$ there.
This is a fairly good approximation also for $T > 1.25\, T_{\rm Ch}$,
but there, an even better approximation is ${\eta'}\approx \eta_{\text{NS}}$,
$M_{\eta'}(T)\approx M_{\text{NS}}(T)$ and ${\eta}\approx \eta_{\text{S}}$,
 $M_{\eta}(T)\approx M_{\text{S}}(T)$.
Finally, when at $T\approx 1.5\, T_{\rm Ch}$ the anomalous mass contribution
 becomes so small that $M_{\text{NS}}(T) = M_{\text{S}}(T)$, 
 Eq. (\ref{M2etas}) enforces anticrossing:
 $M_{\text{NS}}(T)$ and $M_{\text{S}}(T)$ switch, and after this, the 
$\eta$-$\eta'$ complex enters the NS-S asymptotic regime of the vanishing
anomaly influence: $M_{\eta'}(T) \to M_{\text{S}}(T) \to M_{s\bar s}(T)$,
and $M_{\eta}(T) \to M_{\text{NS}}(T) \to M_\pi(T)$, and $\phi(T)\to 0$.

\section{Summary, discussion and conclusions}
\label{summary}

We have studied the temperature dependence of the masses in the $\eta'$-$\eta$ complex
 in the regime of the crossover restoration of chiral and $U_A(1)$ symmetry. We relied 
 on the approach of Ref. \cite{Benic:2014mha},
which demonstrated the soundness of the approximate way in which
the $U_A(1)$-anomaly effects on pseudoscalar masses were introduced and combined
\cite{Klabucar:1997zi,Kekez:2000aw,Kekez:2001ph,Kekez:2005ie,Horvatic:2007qs,Horvatic:2007mi}
with chirally well-behaved DS RLA calculations in order to study $\eta'$ and $\eta$.
 For $T=0$, this was demonstrated \cite{Benic:2014mha} model-independently, with
 only inputs being the experimental values of pion and kaon masses and decay constants,
 and the lattice value of YM topological susceptibility. However, at $T>0$, dynamical
 models are still needed to generate the temperature dependence of non-anomalous
 quantities through DS RLA calculations, and in this paper we use the same chirally
 correct and phenomenologically well-tried model as in numerous earlier $T\geq0$ studies
 ({\it e.g.}, see \cite{Horvatic:2007qs,Benic:2011fv,Horvatic:2010md} and references
 therein).

Presently, we adopt from Ref. \cite{Benic:2014mha} that the anomalous
contribution to the masses is related to the
full QCD topological charge parameter (\ref{defA}), which contains
the massive quark condensates. They give us the chiral crossover
 behavior for high $T$.
 This is crucial, since 
lattice QCD calculations established that for the physical quark masses,
the restoration of the chiral symmetry occurs as a crossover ({\it e.g.,} 
see \cite{Aoki:2012yj,Buchoff:2013nra,Dick:2015twa} and refs. therein)
characterized by the pseudocritical transition temperature $T_{\rm Ch}$.

Nevertheless, what happens with the $U_A(1)$ restoration is still not clear
 \cite{Sharma:2018syt,Fukaya:2017wfq,Burger:2018fvb,Aoki:2012yj}.
Whereas, {\it e.g.}, Ref. \cite{Dick:2015twa} finds its breaking as high as
 $T\sim 1.5\ T_{\rm Ch}$, Ref. \cite{Tomiya:2016jwr} finds that above the
 critical temperature $U_A(1)$ is restored {\it in the chiral limit}, and 
JLQCD collaboration \cite{Fukaya:2017wfq} discusses possible disappearance of the
$U_A(1)$ anomaly and point out the tight connection with the chiral symmetry restoration.
Hence the need to clarify ``{\it if, how (much), and when}" \cite{Aoki:2012yj} 
$U_A(1)$ symmetry is restored. In such a situation, we believe instructive insights
can be found in our study on how an anomaly-generated mass influences the 
$\eta$-$\eta'$ complex, although this study is not on the microscopic level.

Since JLQCD collaboration \cite{Fukaya:2017wfq} has recently stressed that the chiral
symmetry breaking and $U_A(1)$ anomaly are tied for quark bilinear operators, we again
recall how Ref. \cite{Benic:2014mha} provided support for the earlier proposal of Ref.
\cite{Benic:2011fv} relating DChSB to the $U_A(1)$-anomalous mass contributions in 
the $\eta'$-$\eta$ complex.
This adds to the motivation to determine the full QCD topological charge parameter
(\ref{defA}) on lattice from simulations in full QCD with massive, dynamical quarks
[besides the original motivation \cite{Shore:2006mm,Shore:2x} to remove the
 systematic ${\cal O}({1}/{N_c})$ uncertainty of Eq. (\ref{AapproxChiYM})].
More importantly, this ties the $U_A(1)$ symmetry breaking and restoration to the
chiral symmetry ones. It ties them in basically the
same way in the both references \cite{Benic:2011fv} and \cite{Benic:2014mha} (and
here), except that the full QCD topological charge parameter (\ref{defA}) enables the
crossover $U_A(1)$ restoration by allowing the usage of the massive quark condensates.
But, if the chiral condensate ({\it i.e.}, of {\it massless} quarks) is used in 
extending the approach of Ref. \cite{Benic:2014mha} to finite temperatures, 
the $T>0$ results are, in essence, very similar to those
in Ref. \cite{Benic:2011fv}: the quick chiral phase transition leading to 
quick $U_A(1)$ symmetry restoration at $T_{\rm Ch}$ 
(consistently with Ref. \cite{Tomiya:2016jwr}),
which causes not only the empirically supported \cite{CsorgoVertesiSziklaiVargyas}
drop of the $\eta'$ mass, but also an even larger $\eta$ mass drop;
 if $M_{U_A(1)}^2(T) \propto \beta(T) \to 0$ abruptly when $T\to T_{\rm Ch}$,
Eq. (\ref{massMatrix}) mandates  $M_\eta(T\to T_{\rm Ch})\to M_\pi(T_{\rm Ch})$
equally abruptly (as in Ref. \cite{Benic:2011fv}). However, no experimental
 indication for this has ever been seen, although this is a more drastic
 fall than for the $\eta'$-meson.

The present paper predicts a more realistic behavior of $M_\eta(T)$ thanks
to the smooth chiral restoration, which in turn yields the smooth, partial
$U_A(1)$ symmetry restoration (as far as the masses are concerned)
making various actors in the $\eta$-$\eta'$ complex behave quite differently
from the abrupt phase transition (such as that in Ref. \cite{Benic:2011fv}).
In particular, the $\eta$ mass is now not predicted to drop, but to only rise
after $T\approx T_{\rm Ch}$, just like the masses of other (almost-)Goldstone
 pseudoscalars, which are free of the $U_A(1)$ anomaly influence.  Similarly
 to $T=0$, $\eta$ agrees rather well with the $SU(3)$ flavor state $\eta_8$
until the anticrossing temperature, which marks the beginning of the asymptotic
NS-S regime, where the anomalous mass contributions become increasingly negligible
and $\eta \to \eta_{\rm NS}$.

In contrast to $\eta$, the $\eta'$ mass $M_{\eta'}(T)$ does fall
almost as in the case of the sharp phase transition, where its lower limit,
namely $M_{s\bar s}(T)$, is reached at $T_{\rm Ch}$ \cite{Benic:2011fv}.
 Now, $M_{\eta'}(T)$ at its minimum (which is only around $1.13\,T_{\rm Ch}$ 
 because of the rather extended crossover) is some 20 to 30 MeV above
  $M_{s\bar s}(T)$, after which they both start to grow appreciably,
  and $M_{\eta'}(T)$ is reasonably
 approximated by $M_{\eta_0}(T)$ up to the anticrossing. Only beyond the
 anticrossing at $T\approx 1.5\, T_{\rm Ch}$, the effective
 restoration of $U_A(1)$ regarding the $\eta$-$\eta'$ masses occurs,
 in the sense of reaching the asymptotic regime $M_{\eta'}(T)\to M_{s\bar s}(T)$.
 Another, less illustrative qualitatively, but more quantitative criterion
for the degree of $U_A(1)$ restoration is that there, at $T\approx 1.5\, T_{\rm Ch}$,
 $M_{U_A(1)}$ is still slightly above 40\%, and at
 $T\approx 1.8\, T_{\rm Ch}$ still around 14\% of its $T=0$ value.
 Thus, the drop to the minimum of $M_{\eta'}(T)$ around $1.13\,T_{\rm Ch}$
 in any case signals only a partial $U_A(1)$ restoration.

This $M_{\eta'}(T)$ drop is around 250 MeV, which is consistent with
the present empirical evidence claiming that it is at least 200 MeV
 \cite{CsorgoVertesiSziklaiVargyas}.
For comparison with some other approaches exploring the interplay of the
chiral phase transition and axial anomaly, note that the $\eta'$ mass drop
around 150 MeV is found in the functional renormalization group approach
\cite{Mitter:2013fxa}. A very recent analysis within the framework of the
$U(3)$ chiral perturbation theory found that the (small) increase of the
masses of $\pi$, $K$ and $\eta$ after around $T\sim 120$ MeV, is accompanied
by the drop of the $\eta'$ mass, but only by some 15 MeV \cite{Gu:2018swy}.

Admittedly, the crossover transition leaves more space for model dependence,
since some model changes which would make the crossover even smoother would
reduce our $\eta'$ mass drop. Nevertheless, there are also changes which would
make it steeper, and those may, for example, help $M_{\eta'}(T)$ saturate
 the $M_{s\bar s}(T)$ limit.
Exploring such model dependences, as well as attempts to further reduce
them at $T>0$ by including more lattice QCD results, must be relegated to
the future work.
However, already here we can note a motivation for varying the presently
isosymmetric {\it model} current $u$- and $d$-quark mass of 5.49 MeV.
Since it is essentially a phenomenological model parameter, it cannot be
quite unambiguously and precisely related to the somewhat lower PDG values
$m_u=2.2^{+0.5}_{-0.4}$ MeV and $m_d=4.70^{+0.5}_{-0.3}$ MeV \cite{Tanabashi:2018oca}.
Still,  
their ratio $m_u/m_d=0.48^{+0.07}_{-0.08}$ is quite instructive in the present
context, since the QCD topological susceptibility $\chi$ (\ref{chiShore_small_m})
and charge parameter $A$ (\ref{defA}) contain the current quark masses in the
form of harmonic averages of $m_q\, \langle {\bar q}q\rangle$ ($q=u,d,s$).
Since a harmonic average is dominated by its smallest argument, our
$\chi$ (\ref{chiShore_small_m}) and $A$ (\ref{defA}) are  dominated by the
lightest flavor, providing the motivation to venture beyond the precision
of the isospin limit and in the future work explore the maximal isospin
violation scenario \cite{Kharzeev:1998kz} within the present treatment
of the $\eta$-$\eta'$ complex.

\vskip 3mm 
\noindent {\bf Acknowledgment:}
This work was supported in part by the Croatian Science Foundation under
the project number 8799, and by
 STSM grants from COST Actions CA15213 THOR and CA16214 PHAROS.
D. Kl. thanks for many helpful discussions with T. Cs\"org\H{o}
 and D. Blaschke.



\end{document}